\documentclass[conference]{IEEEtran}
\usepackage{subfig}
\usepackage{array}
\usepackage{graphicx}
\usepackage{amsmath}
\usepackage{algorithmic}
\usepackage{algorithm}
\usepackage[bookmarks=false]{hyperref}
\usepackage{doi}
\usepackage{url}

\usepackage{setspace}
\usepackage{enumitem}
\usepackage{graphicx,lipsum}
\usepackage{caption}
\usepackage{tikz}

\newcolumntype{P}[1]{>{\centering\arraybackslash}p{#1}} 
\newcolumntype{M}[1]{>{\centering\arraybackslash}m{#1}} 
\newcommand\copyrighttext{%
  \footnotesize \copyright~2015 IEEE. Personal use of this material is permitted. Permission from IEEE must be obtained for all other uses, in any current or future media, including reprinting/republishing this material for advertising or promotional purposes, creating new collective works, for resale or redistribution to servers or lists, or reuse of any copyrighted component of this work in other works.}
\newcommand\copyrightnotice{%
\begin{tikzpicture}[remember picture,overlay]
\node[anchor=south,yshift=10pt] at (current page.south) {\fbox{\parbox{\dimexpr\textwidth-\fboxsep-\fboxrule\relax}{\copyrighttext}}};
\end{tikzpicture}%
}
\begin{document}
\bibliographystyle{IEEEtran}
%

\title{Radio Co-location Aware Channel Assignments for Interference Mitigation in Wireless Mesh Networks}

\author{\IEEEauthorblockN{Srikant Manas Kala, M Pavan Kumar Reddy, Ranadheer Musham, and Bheemarjuna Reddy Tamma}
\IEEEauthorblockA{ Indian Institute of Technology Hyderabad, India\\
Email: [cs12m1012, cs12b1025, cs12b1026, tbr]@iith.ac.in}}
%


\maketitle
\copyrightnotice
\begin{abstract}
Designing high performance channel assignment schemes to harness the potential of multi-radio multi-channel deployments in wireless mesh networks (WMNs) is an active research domain. A pragmatic channel assignment approach strives to maximize network capacity by restraining the endemic interference and mitigating its adverse impact on network performance. Interference prevalent in WMNs is multi-faceted, \textit{radio co-location interference} (RCI) being a crucial aspect that is seldom addressed in research endeavors. In this effort, we propose a set of intelligent channel assignment algorithms, which focus primarily on alleviating the RCI. These graph theoretic schemes are structurally inspired by the spatio-statistical characteristics of interference. We present the theoretical design foundations for each of the proposed algorithms, and demonstrate their potential to significantly enhance network capacity in comparison to some well-known existing schemes. We also demonstrate the adverse impact of radio co-
location interference on the network, and the efficacy of the proposed schemes in successfully mitigating it. The experimental results to validate the proposed theoretical notions were obtained by running an exhaustive set of ns-3 simulations in IEEE 802.11g/n environments.
\end{abstract}
\section{Introduction}

Wireless Mesh Networks (WMNs) are a precursor to the next-generation of wireless communication systems, which envision a seamless network-convergence through integration of cellular mobile technologies \emph{viz.}, 4G/LTE and 5G, with IEEE 802.11 wireless local area networks (WLANs), on a single delivery platform \cite{12Capone}. 
The benefits of increased throughput, enhanced connectivity and reduced latency that are promised by a multi-radio multi-channel (MRMC) WMN depend upon the intensity of prevalent interference. Minimizing the interference in wireless networks is thus a fundamental system design objective. The most effective tool of achieving this goal is an optimal and feasible channel assignment (CA) to the radios in a WMN. A prudent CA will reign in the interference and its detrimental effects thereby improving overall network performance. Thus, interference alleviation is not a characteristic of the network itself, but instead of the CA scheme.

\section{Motivation and Related Research Work}
In a recent study, we have focused on \textit{radio co-location interference} (RCI) \cite{Manas} and its detrimental impact on WMN performance, a problem that has so far remained largely unaddressed. RCI is caused and experienced by spatially co-located radios (SCRs) at a node that have been allocated identical channels to communicate on. We propose an \textit{Enhanced Multi-Radio Multi-Channel Conflict Graph} (E-MMCG) model, which enables a CA algorithm to mitigate the adverse impact of RCI and substantially enhances CA performance. E-MMCG model takes into account the interference scenarios spawned and experienced by SCRs at a node in the WMN, and adequately represents them in its conflict graph (CG). E-MMCG model accomplishes this by adding an edge between two vertices of the CG \textit{iff} the respective wireless links in the WMN emanate from, or terminate at, the same node and have been assigned an identical channel. The E-MMCG thus generated is a comprehensive representation of all possible conflict 
links in a WMN, including the RCI scenarios, and serves as an ideal input to CA schemes.\\
The CA problem is an NP-Hard problem \cite{NPcomplete}. The CA approaches can be broadly classified into three categories, \emph{viz.} static, dynamic and hybrid. A static scheme mandates a fixed radio-channel mapping throughout the session for which the network is operational. Dynamic CA requires updating of the CA continuously, based on the analysis of network performance metrics. Hybrid CAs employ characteristics of both static and dynamic CA schemes, efficiently implementing a fixed CA on some radio interfaces and a dynamic CA on others. Several CA schemes have been proposed in earlier research studies \cite{Ding}. Yet, to the best of our knowledge, none of the proposed CA schemes incorporates RCI alleviation as a design objective. We now mention a few relevant CA schemes. A centralized Breadth First Traversal (BFT) approach BFS-CA, has been suggested in  \cite{22Ramachandran} where channels are assigned to nodes, after performing a BFT with the Gateway node as the reference. A static maximum clique 
based algorithm is discussed in \cite{17Xutao}. In \cite{24Aizaz} authors proposed MaIS-CA, a Maximum Independent Set (MIS) based high performance CA, where MIS of the \textit{conflict graph} (CG) is determined iteratively, and channels are assigned to the nodes of the MIS in each step. A hybrid CA scheme is proposed in \cite{LLLA} which adopts a link-layer and learning automata approach. In \cite{Gravity}, authors propose an enhanced version of gravitational search algorithm (IGSA) by combining it with a local operator. IGSA CA offers improved network performance and also ensures network connectivity. Authors propose a CA scheme in \cite{Part} that makes use of partially overlapping channels to optimize spectrum usage while guaranteeing minimal impact of interference. A link-centric CA scheme which employs physical interference model and considers the number of radios as a constraint is proposed in \cite{Linkc}. It is a topology preserving scheme that considers the  derived upper and lower bounds of the 
prevalent interference while assigning channels to links. A joint channel allocation approach is described in \cite{Joint} which employs an optimal power control model. This model considers parameters such as the interference present in a WMN and power consumption of the WMN nodes. A centralized static CA scheme is proposed in \cite{23Cheng}, where nodes of the CG are assigned channels so that the \textit{total interference degree} (TID) decreases, resulting in an efficient CA. TID is a theoretical estimate which gives a measure of the intensity of interference prevalent in a WMN. It is the sum of individual \textit{conflict numbers} of all links, where conflict number of a link represents the total number of potential link conflicts experienced by that particular link in a WMN.\\
However, the absence of RCI mitigation as a design consideration is bound to hamper the efficiency of a CA, since it does not restrain the RCI. It is imperative that CA algorithms be \textit{Radio Co-location Aware} or RCA \emph{i.e.}, ensure RCI mitigation while assigning channels to the radios. In the current work, we propose two RCA CA algorithms. We opt to design static CA schemes, as the primary bottlenecks in dynamic and hybrid CAs are the channel switching delays, and the need for co-ordination between radios to ensure that they are on the same channel when they intend to communicate \cite{8Raniwala}.

\section{Problem Defintion}
Let $G=(V,E)$ represent an arbitrary MRMC WMN comprising of $n$ nodes, where $V$ denotes the set of all nodes in the WMN and $E$ denotes the set of wireless links between nodes which lie within each other's transmission range. Each node $i$ is equipped with a random number of identical radios $R_i$, and is assigned a list of channels $C_i$. The number of available channels is greater than the maximum number of radios installed on any node in the WMN. For the WMN described above, we propose \textit{RCA} CA schemes, which allocate channels to radios of every node $i$ in the WMN \emph{i.e.}, $C_i = CA_{RCA}(G)$, so as to efficiently mitigate the detrimental impact of RCI. 
  
\section{Radio Co-location Aware Channel Assignments}
We now present the two RCA CA algorithms. The proposed algorithms benefit from the following design considerations.
\subsubsection{Enhanced Conflict Graph Model}
Proposed RCA CA algorithms employ E-MMCG model \cite{Manas} to generate the E-MMCG of a WMN, which serves as input. The broad-based E-MMCG model accounts for the prevalent RCI and adequately represents it in the CG of the WMN.
\subsubsection{Radio Co-location Optimization (RCO)}
This is the signature functionality of the RCA CA algorithms. It mitigates RCI by ensuring that SCRs are not allocated identical channels to communicate on. It also serves as an optimization step by restraining the detrimental effect of interference over network performance.
\subsubsection{Spatio-Statistical Interference Alleviation}
 Due to inherent rigidity of radio-channel mapping in a static CA, it can only address the spatio-statistical aspects of the three dimensional interference-mitigation problem, the third dimension being the temporal or dynamic characteristics \cite{Manas3}. A CA scheme primarily catering to the spatial features of interference will ensure that links operating on overlapping channels are efficiently interspersed with links operating on orthogonal channels. An intelligent spatio-statistical scheme, in addition to spatial prudence, will strive to evenly distribute the available channels among the radios, thereby facilitating an improved CA with enhanced fairness.
 \subsubsection{Network Topology Preservation} 
Preserving network topology at the cost of increased interference holds its relevance in reduced propagation delays between end-nodes, and uninterrupted connectivity to the end-user. Further, a CA approach should not alter the original WMN topology to ensure the functional independence of the physical layout of a WMN from the CA exercise. RCA CAs guarantee that the WMN topology is preserved after CA deployment. \\
Thus, the proposed algorithms are graph theoretic approaches which restrain RCI by employing the twin interference mitigation features, \emph{viz.}, the E-MMCG model and RCO. Further, they incorporate spatial and statistical dimensions of the prevalent interference in their structural design, to fashion high-performance RCA CAs. Notations common to both the algorithms are, $G$: The WMN graph, $G_c$: E-MMCG of the WMN graph, $CS$ : The set of $M$ available channels,  $Adj_i$ : The set of nodes adjacent to node $i$ in $G$, $Ch_i$  : The list of channels allocated to the radios of node $i$ in $G$; $Ch_i$ may have duplicate elements, which will reflect the impact of RCI at node $i$. $TID(G)$ : The function which computes the TID.

\subsection{RCA Optimized Independent Set (OIS) CA}
This graph-theoretic RCA CA scheme appeals to the statistical aspects of channel assignment. We contend that given two CA schemes with similar spatial patterns, the one with a more proportionate distribution of available channels among the radios of a WMN will perform better. 

The method of assigning channels to radios based on the maximal independent set (MIS) approach finds numerous references in the research literature \cite{yutao} \cite{24Aizaz}. However, the MIS approach is not statistically pragmatic with respect to the channel allocation exercise, and an independent set (IS) approach stands to fare better. For example, the MaIS-CA scheme elucidated in \cite{24Aizaz} suggests that in each iteration, an MIS of the updated CG is determined, all vertex elements of the MIS are assigned a common channel and then removed from the CG. An MIS can not add even a single vertex to its element set, lest it violate its independence. Further, an MIS may or may not be a \textit{maximum} IS. Yet, we can infer that as cardinality of vertex set of the CG decreases in subsequent iterations, the MaIS-CA scheme is disposed to generate MISs with decreasing cardinalities as well. Since in each step, a channel is assigned to all the vertices of an MIS, the distribution of channels among radios is 
bound to be uneven. To validate our argument we employ MaIS-CA scheme to generate CAs for WMN grids of size  $(N\times N)$, where $N\in \{5,...,9\}$. Every node is equipped with two identical radios and three orthogonal channels $C_1, \ C_2 \ \& \ C_3$, are available. $C_1$ is the default channel initially assigned to all radios to generate the E-MMCG. We ascertain the statistical evenness of  MaIS-CA in various WMN grid layouts by determining the ratio of the number of radios operating on each channel denoted by $R_{C1}$, $R_{C2}$ $\&$ $R_{C3}$, normalized by the smallest value among the three. The results are illustrated in Table~\ref{MvsO}. It can be discerned that there is a skewed distribution of channels among the radios. The number of radios which are allocated $C_3$ is always at least $50\%$ more than those on $C_1$, which is the default channel and is consistently under utilized in the final CA. The difference between $R_{C2}$ and $R_{C1}$ is also never below $33\%$, highlighting a statistically 
uneven allotment of channels to radios.
\begin{table} [h!]
\caption{Channel distribution, RCA OIS-CA vs MaIS-CA}
\tabcolsep=0.11cm
\begin{tabular}{|M{1.5cm}|M{1.5cm}|M{2.5cm}|M{2.5cm}|}
\hline 
    \multicolumn{1}{|c|}{\textbf{Grid}}&\multicolumn{1}{|c|}{\textbf{Num of}}&\multicolumn{2}{|c|}{\textbf{R$_{C1}$ : R$_{C2}$ : R$_{C3}$}}\\ \cline{3-4}
     \multicolumn{1}{|c|}{\textbf{Size}}&\multicolumn{1}{|c|}{\textbf{Radios}}&\textbf{MaIS}&\textbf{RCA OIS}\\
\hline  
5$\times$5&50&1.00 : 1.63 : 1.94&1.00 : 1.06 : 1.06\\
\hline  
6$\times$6&72&1.00 : 1.33 : 1.66&1.00 : 1.09 : 1.33\\
\hline  
7$\times$7&98&1.00 : 1.56 : 1.69&1.00 : 1.00 : 1.16\\
\hline
8$\times$8&128&1.00 : 1.48 : 1.64&1.00 : 1.00 : 1.28\\
\hline
9$\times$9&162&1.00 : 1.58 : 1.57&1.08 : 1.00 : 1.29\\
\hline  
\end{tabular} 
\label{MvsO}
\end{table}

\renewcommand{\algorithmicrequire}{\textbf{Input:}}
\renewcommand{\algorithmicensure}{\textbf{Output:}}
\begin{algorithm}[htb!] 
\caption{RCA Optimised Independent Set CA}
\label{OIS}
\begin{algorithmic}[1]
{\fontsize{9}{10}
\REQUIRE $G = (V,E)$, $G_c = (V_c,E_c)$, $CS =\{1, 2,...M\}$ \\
\ENSURE RCA OIS Channel Assignment For $G$ \\
\line(1,0){236}
\STATE $IS \leftarrow FindIndependentSets(G_c)$. \COMMENT {$IS$ : Set of mutually exclusive Independent Sets of vertices of $G_c$.}
\STATE	$Channel \in CS$, $Channel \leftarrow 1$.
\FOR {$IndSet \in IS$}
\FOR {$Node \in IndSet$}
\STATE $Node \leftarrow Channel$
\ENDFOR
\STATE $Channel \leftarrow Channel \% M + 1$
\ENDFOR\\
\COMMENT{ Let $V_r$ be the subset of all vertices in $V_c$ which denote a link emanating from a particular radio $r$ in $G$. Let $C_{last}$ be the $Channel$ assigned to the last element processed in $V_r$.}
\STATE 	$r \leftarrow C_{last}$   \COMMENT{Facilitates improvised vertex coloring.}
\FOR {$i \in V$} 
\STATE $Num_i \leftarrow i$ \COMMENT{Number the nodes from $(1,...,N)$.} 
\STATE Determine $Ch_i$ and $Adj_i$
\ENDFOR \\
\COMMENT {Ensure Topology Preservation in $G$.}
\FOR {$i \in V$}
\FOR {$j \in Adj_{i}$} 
\IF {$((Num_i < Num_j)$  $\&\&$ $(\lvert Ch_{i} \cap Ch_{j}\lvert$ $==0))$}
\STATE 	$Ch_{j}  \leftarrow Ch_{j} + \{c_{com}\} - \{c_{dif}\}$ $\lvert \ \{(c_{com}\in Ch_{i})$ $\&\&$ $(c_{dif} \in Ch_{j})$  $\&\&$ $(TID(G)$ \textit{is minimum)}$\}$
\ENDIF
\ENDFOR
\ENDFOR
\STATE \textit{Perform Radio Co-location Optimization in $G$} \COMMENT{Steps described in Algorithm~\ref{radcol}.}
}
\end{algorithmic}
\end{algorithm}

We now present the RCA \textit{Optimized Independent Set} (OIS) CA algorithm in Algorithm~\ref{OIS}. Here, we approach the CA problem as an improvised vertex-coloring problem, ensuring that no radio on any node in the original WMN graph $G$ is assigned multiple channels. For a smooth discourse, let us consider an arbitrary radio $r$ in the WMN. All the wireless links emanating from $r$ will be represented in the E-MMCG $G_c$ by a subset of vertices $V_r$, which will form a \textit{clique} as every vertex element in $V_r$ will be connected to every other vertex in the subset. The initial step is to partition the vertex set of the E-MMCG into ISs. This is done by traversing each node exactly once, and allotting it an IS. If a node can not be assigned to any of the existing ISs, a new IS is created of which it becomes the first node. If a node can be assigned to more than one ISs, the IS with minimum cardinality is chosen. Next, all nodes of an IS are assigned the same channel, and the channel to be assigned to 
the next IS is 
selected in a cyclic fashion. But any two vertices in the clique $V_r$ will never be a part of the same IS, as they are pair-wise adjacent to each other in the E-MMCG. This fact forbids translating the CA problem into a simple vertex coloring problem. To overcome this handicap, we adopt a \textit{selective} vertex coloring approach. As all the vertices in $V_r$ lie in different ISs, depending upon the number of available channels and the cyclic channel assignment, they may have been assigned different channels. Therefore, after generating all the ISs we identify the color (channel) that has been assigned to the elements of $V_r$ the maximum number of times, and assign radio $r$ in $G$ that particular channel. If two or more colors (channels) have been assigned an equal number of times or all elements in $V_r$ have been assigned different colors, we pick a color (channel) randomly and assign it to $r$. The insight behind this improvisation is that a vertex in E-MMCG represents a link between two radios, and 
allocating radio $r$ a channel that has been assigned to maximum number of vertices in $V_r$ will ensure increased connectivity in the WMN graph. From a computational perspective, the cyclic manner of assigning channels to ISs facilitates an $O(1)$ time implementation of the selective vertex coloring approach. The underlying idea is that the color (channel) assigned to the last element in $V_r$ denoted by $C_{last}$, will always be one of the maximally assigned channels among the vertices of  $V_r$.\\
Despite its design which enhances connectivity in the WMN, the initial OIS-CA may disrupt the original topology and may even lead to a disconnected WMN graph. It is of great importance to preserve the original WMN topology which reflects the intended physical span of the wireless network. However, designing an optimal topology preserving CA algorithm is an established NP-hard problem \cite{TopoNPHard}. Hence, the algorithm described in steps $10-20$ of Algorithm~\ref{OIS} is a smart heuristic approach specifically tailored to ensure topology preservation after the initial channel assignment. The algorithm first sequentially orders the nodes of the WMN graph $G$, and assigns every node $i$ a number $Num_i$. For each node $i$ in $G$, it then determines the sets ($Ch_i$) and ($Adj_i$). Further, for every $i$ in $G$, every neighboring node $j$ in $Adj_i$ such that $Num_i<Num_j$ is scanned to determine if $i$ $\&$ $j$ share a common channel to communicate. If not, then WMN topology is violated and a \textit{
forward correction} technique is adopted to establish a connection between the two nodes and to restore the topology. The constraint $Num_i<Num_j$ ensures that in every topology preserving channel re-assignment, a channel from node $i$ is picked and assigned to node $j$ \emph{i.e.}, a correction happens only in the forward direction. This mechanism rules out the possibility of a backward link disruption on a node while attempting to re-establish its connection on another link. Further, the choice of the $(c_{com},c_{dif})$ pair from $Ch_{i}$ and $Ch_{j}$, respectively, depends upon the configuration with minimum TID value \emph{i.e.}, the maximum decrease in prevalent interference.\\
The final and most crucial step involves RCI mitigation. It is a performance enhancement feature which combines E-MMCG model for initial channel assignment with post channel allocation radio co-location optimization. Thus, the twin methods of restraining RCI have been employed in conjugation.
\renewcommand{\algorithmicrequire}{\textbf{Input:}}
\renewcommand{\algorithmicensure}{\textbf{Output:}}
\begin{algorithm}[htb!] 
\caption{Radio Co-location Optimization}
\label{radcol}
\begin{algorithmic}[1]
{\fontsize{9}{10}
\REQUIRE$G = (V,E)$, $CS =\{1, 2,...M\}$ \\
\ENSURE Mitigate RCI in $G$ $\&$ minimize $TID(G)$ \\
\line(1,0){236}
\FOR {$i \in V$}
\STATE Determine $Ch_i$ and $Adj_i$ 
\ENDFOR
\FOR {$ i \in V$}
\IF  {$(((j_1,j_2,..,j_k,Channel) \in Ch_{i})$ \ $\&\&$ \ $(j_1=j_2=...=j_k=Channel)$ $\&\&$ \ $( k \geq 2) ))$}
\STATE $j_1\leftarrow Channel, \ j_2\leftarrow c_2,..., \ j_k \leftarrow c_k\lvert\{((c_2,...,c_k)\in CS$ \textit{are distinct if possible)} $\&\&$ $(TID(G)$ \textit{is minimum}$)\}$
\ENDIF
\ENDFOR
\FOR {$i \in V$}
\FOR {$j \in Adj_i$}
\STATE $Get$ $c_{ij}$ \COMMENT {Current channel of wireless link $(i,j)$}
\FOR   {$c_{dif} \in CS$}
\IF	{$((c_{ij} \leftarrow  c_{dif})$ $\&\&$ $(NetTopoPreserved())$ $\&\&$ $(TID(G)$ $decreases))$}
\STATE 	$Ch_{i}  \leftarrow Ch_{i} + \{c_{dif}\} - \{c_{ij}\}$ 
\STATE 	$Ch_{j}  \leftarrow Ch_{j} + \{c_{dif}\} - \{c_{ij}\}$
\ENDIF
\ENDFOR
\ENDFOR
\ENDFOR
}
\end{algorithmic}
\end{algorithm}

\begin{figure}
  \centering%
  \begin{tabular}{cc}
  \subfloat[TID estimates for grid WMNs]{\includegraphics[width=.55\linewidth, height=4cm]{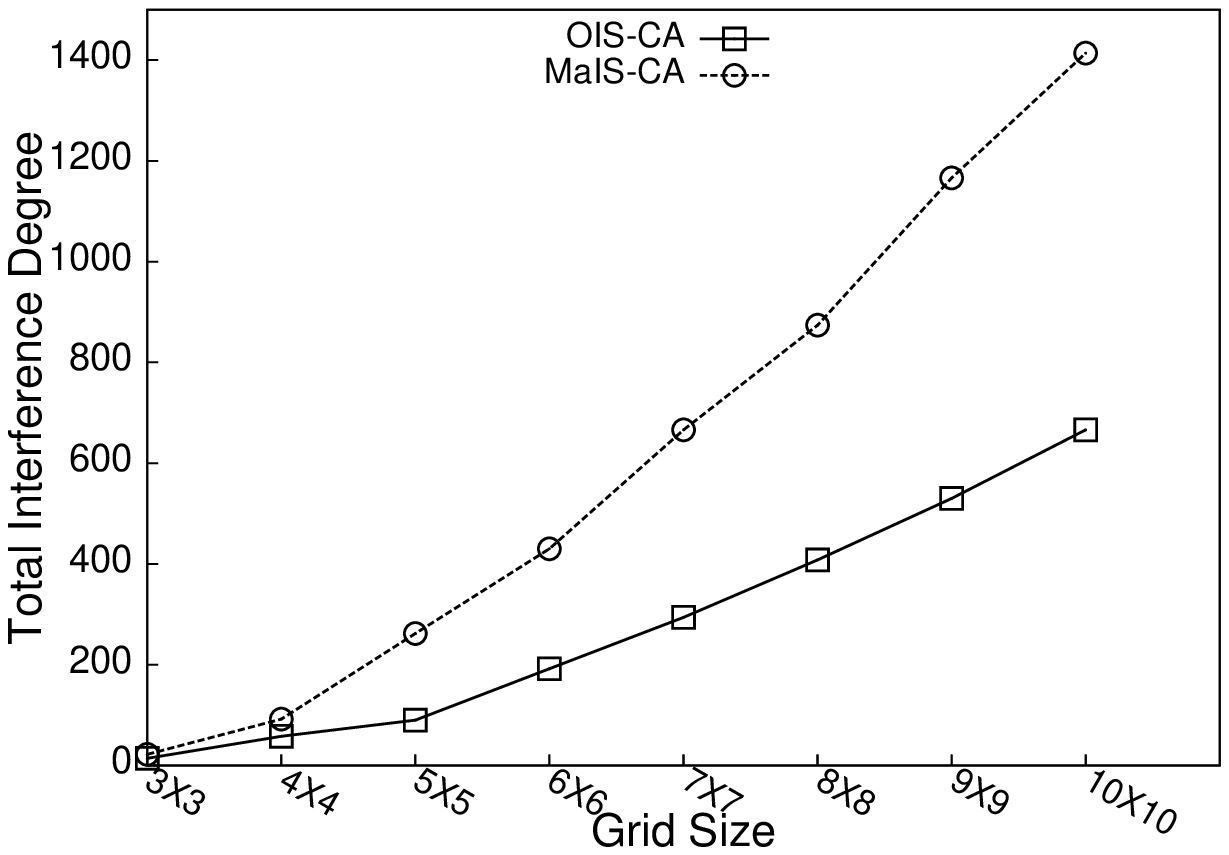}}  \hfill%
   \subfloat[A sample $5 \times 5$ grid]{\includegraphics[width=.4\linewidth, height=4cm]{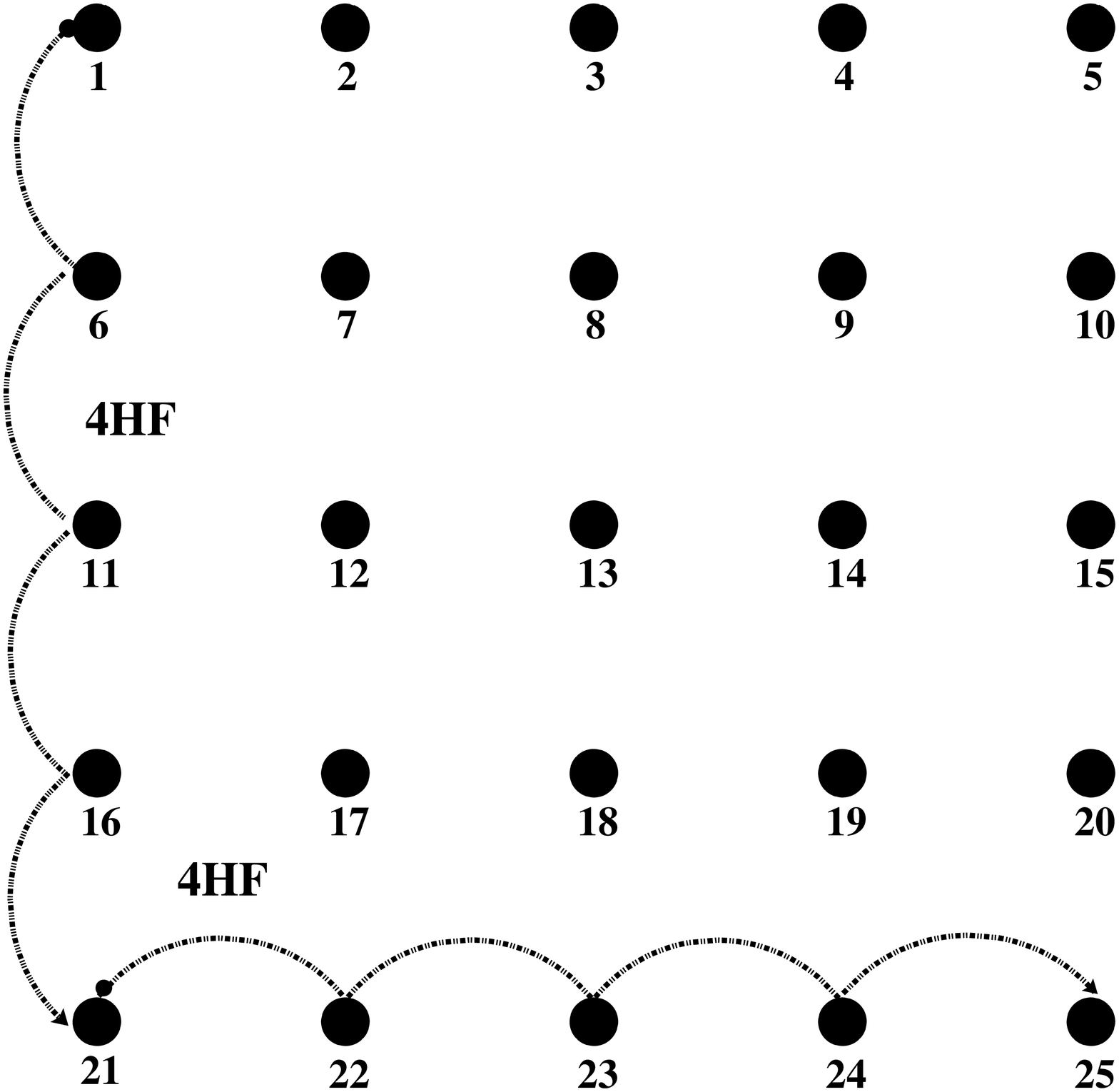}}%
    \end{tabular}
    \caption{TID values of RCA OIS-CA vs MaIS-CA}
     \label{TID}
\end{figure}

RCO function presented in Algorithm~\ref{radcol} commences by ascertaining $Ch_i$ and $Adj_i$ for each node $i$ of $G$. It explores $Ch_i$ of each node to determine if two or more SCRs have been assigned an identical channel to operate on. If so, it attempts to alleviate RCI by re-assigning all such SCRs barring one, distinct channels from the set of all available channels $(CS)$, ensuring that the final configuration results in a minimum TID value. One of the SCRs continues to operate on the channel it was originally assigned, so as not to disrupt an existing wireless link of the node in context. After performing RCI mitigation, the function analyzes each wireless link in the WMN exactly once. This is done to check if the current channel assigned to a link can be replaced with an alternate channel $(c_{dif})$ from $CS$ so that the overall TID estimate decreases, with an important caveat that the underlying WMN topology is preserved denoted by function $(NetTopoPreserved())$. If a desirable channel 
replacement for a link is found, the current channel $(c_{ij})$ connecting the nodes  $i$ \& $j$ is removed from both $Ch_i$ \& $Ch_j$, and the preferred channel  $(c_{dif})$ is added to both.\\
The intelligent features described above, coupled with the characteristic that the ISs are generated concurrently in OIS-CA, result in ISs of comparable cardinalities. In sharp contrast, cardinalities of MISs generated consecutively in MaIS-CA fail to provide for a balanced distribution of channels among radios. Table~\ref{MvsO} illustrates the performance of RCA OIS CAs in terms of statistical evenness. OIS-CA outperforms MaIS-CA and the following two aspects elicit this observation. Firstly, in OIS-CA the difference between the cardinalities of any two sets of radios among $R_{C1}$, $R_{C2}$ $\&$ $R_{C3}$, never exceeds $35\%$ and the difference between at least a pair of mentioned sets is always under $10\%$. This ensures a balanced channel assignment. Secondly, the default channel $C_1$ is not under utilized and is assigned to almost as many radios as channel $C_2$, guaranteeing fairness in spectrum utilization. The relevance of judicious statistical distribution and fairness in channel utilization 
demonstrated by 
RCA OIS-CA, is substantiated by the TID estimates depicted in Figure~\ref{TID} (a). The grid WMNs are of size $(N\times N)$, where $N\in \{3,...,10\}$, each node is equipped with 2 radios and 3 orthogonal channels are available. It can be inferred from the consistently lower TID estimates exhibited by RCA OIS-CA as compared to MaIS-CA, that it is the more efficient CA scheme owing to its intelligent algorithm design. This theoretical conclusion will be substantiated by the experimental results presented in later sections.
%
        
\subsection{Elevated Interference Zone Mitigation (EIZM) CA }
We now propose the RCA \textit{EIZM} algorithm which is primarily pivoted on the spatial characteristics of endemic radio interference, but also factors in the statistical dimension. The motivating principle catalyzing the algorithm design is that the intensity of prevalent interference fluctuates within a wireless network, causing localized pockets of unusually high interference levels. We call wireless links which form the epicenter of severe performance bottlenecks in the WMN as \textit{Elevated Interference Zones} (EIZs). Taking a cue from the study on impact of interference on a WMN in \cite{IntImpWMN}, we offer the argument that EIZs impede network performance drastically as they substantially degrade the \textit{signal to noise plus interference ratio} (SINR) on neighboring links in the network. A WMN with a large number of links on which data transmissions experience strong levels of interference will lead to a dismal aggregate network capacity, even though the communication over the remaining links 
exhibits a high SINR. This is due to the inherent multi-hop nature of packet exchange among nodes. In contrast, a WMN in which the average SINR of the network is same as that of the former, but all radio links experience comparable levels of interference, will perform better than the former. 
Extreme fluctuations in interference levels impede network performance. Thus, an EIZ is a particular wireless link in the WMN that may cause severe degradation of the network capacity. The question now remains as to how do we precisely characterize an EIZ for an accurate physical modeling? We consider a wireless link with high number of adjacent links as a potential EIZ, the extent of whose severity is predicated upon the channels allocated to it and its neighboring links. The intuitive notion is that the channel assignment on an EIZ tends to have an enormous impact on the network performance due to its high number of potential conflict links. A favorable channel assigned to an EIZ may significantly improve the overall network capacity while an ill-chosen channel may exacerbate the effects of interference manifold. We adopt a view that eroding the detrimental impact of interference on an EIZ will occasion a ripple effect that will certainly reduce the adverse effects of interference on its adjacent links.\\
We now translate our theoretical proposition into a feasible practical implementation \emph{i.e.,} correlate the concept of an EIZ in a WMN with its representation in the E-MMCG. An EIZ in the WMN is identified by its corresponding node in the E-MMCG, and by labeling E-MMCG nodes as EIZs, the algorithm pinpoints the respective performance bottleneck links in the WMN. The RCA EIZM-CA algorithm accepts the E-MMCG of a WMN as input, and considers EIZs to be the nodes with the \textit{maximal degree} and nodes which share the \textit{maximal number of neighbors} with an existing EIZ. The maximal degree node signifies a link with the maximum link-adjacency in a WMN and an obvious reference EIZ node to begin with. The subsequently chosen EIZ nodes are those which share the most number of mutual neighbors with the existing EIZ node. The process of EIZ selection should ideally be based on the SINR values of nodes. However, SINR is a temporal or dynamic link quality parameter and can be ascertained only during active 
data transmissions. Thus, the improvised EIZ selection approach considers the theoretical TID estimates as an approximate measure of the impact of interference instead of the dynamic SINR values. The nodes with high TIDs are the first candidates to be labeled EIZs. Further, the scheme intelligently assigns channels to the EIZ nodes so that the detrimental impact of interference on the WMN, represented by the TID estimate, reduces. This is a spatial algorithm design strategy wherein the intense EIZs \emph{i.e.,} the maximal degree nodes in E-MMCG are efficiently assigned channels first. Thereafter the nodes which share the maximal number of mutual neighbors with them are visited and so forth, triggering an \textit{outward ripple} of interference mitigation.

EIZM approach elucidated in Algorithm~\ref{MaNI}, employs a Breadth First Traversal (BFT) starting from a reference node, which could be a node representing a link to the Gateway or a specifically chosen node. We consider a maximal degree node $v_{maximal}$, which is unarguably an EIZ to be the reference. A \textit{Level Structure} (LS) is generated, which is a partition of the vertex set $V_c$ into subsets of vertexes that lie in the same level of the BFT \emph{i.e.}, vertices which are situated the same number of hop-counts away from $v_{maximal}$ are placed in the same level. We term these subsets of the LS as \textit{level-sets}. Next, to fashion a fair spectrum utilization, each level-set is assigned a channel chosen from CS  in a cyclic manner. This step caters to the statistical aspects of interference alleviation, by attempting to maintain an equitable distribution of channels across radios in the WMN. In addition, vertices of level-sets which differ by a single level are assigned non-overlapping 
channels, thereby substantially reducing link conflicts.

\renewcommand{\algorithmicrequire}{\textbf{Input:}}
\renewcommand{\algorithmicensure}{\textbf{Output:}}
\begin{algorithm}[htb!] 
\caption{RCA Elevated Interference Zone Mitigation CA}
\label{MaNI}
\begin{algorithmic}[1]
{\fontsize{9}{10}
\REQUIRE $G = (V,E)$, $G_c = (V_c,E_c)$, $CS =\{1, 2,...M\}$ \\
\ENSURE RCA EIZM Channel Assignment For $G$ \\
\line(1,0){236}
\STATE Let $v_{maximal} \leftarrow maximalDegNode(V_c)$
\STATE $LS \leftarrow LevelStructure(v_{maximal}, G_c)$  
\COMMENT {$LS$ : Set of level-sets generated by Breadth First Traversal of vertices of $G_c$}
\STATE	$Channel \in CS$, $Channel \leftarrow 1$
\FOR	{$LevSet \in LS$ }
\FOR	{$Node \in LevSet$}
\STATE	$Node \leftarrow Channel$
\ENDFOR
\STATE	$Channel \leftarrow Channel \% M + 1$
\ENDFOR
\FOR	{$LevSet \in LS$}
\STATE	$PrevNode_{EIZ} \leftarrow 0$ 
\FOR 	{$ Node \in LevSet$}
\IF	{$(PrevNode_{EIZ} \leftarrow 0)$}
\STATE	Let $Node_{EIZ} \leftarrow maximalDegNode(LevSet)$ \COMMENT{Maximal degree node in the Level Set is the initial EIZ node}
\ELSE 
\STATE Let $Node_{EIZ}$ $\leftarrow$ \\ $MaximalMutualNeighbor(LevSet,PrevNode_{EIZ})$
\ENDIF
\STATE	  $Node_{EIZ} \leftarrow Channel$, \textit{such that} $TID(G)$ \textit{is minimum}
\STATE    $PrevNode_{EIZ} \leftarrow Node_{EIZ}$
\STATE 	  $LevSet \leftarrow LevSet - \{Node_{EIZ}\}$
\ENDFOR
\ENDFOR
\STATE \textit{Ensure Topology Preservation in $G$} \COMMENT{Perform steps $10$ to $20$ described in Algorithm~\ref{OIS}}
\STATE \textit{Perform Radio Co-location Optimization in $G$} \COMMENT{Steps described in Algorithm~\ref{radcol}}
}
\end{algorithmic}
\end{algorithm}

\begin{figure}
  \centering%
  \begin{tabular}{cc}
   \subfloat[A sample E-MMCG input]{\includegraphics[width=.5\linewidth]{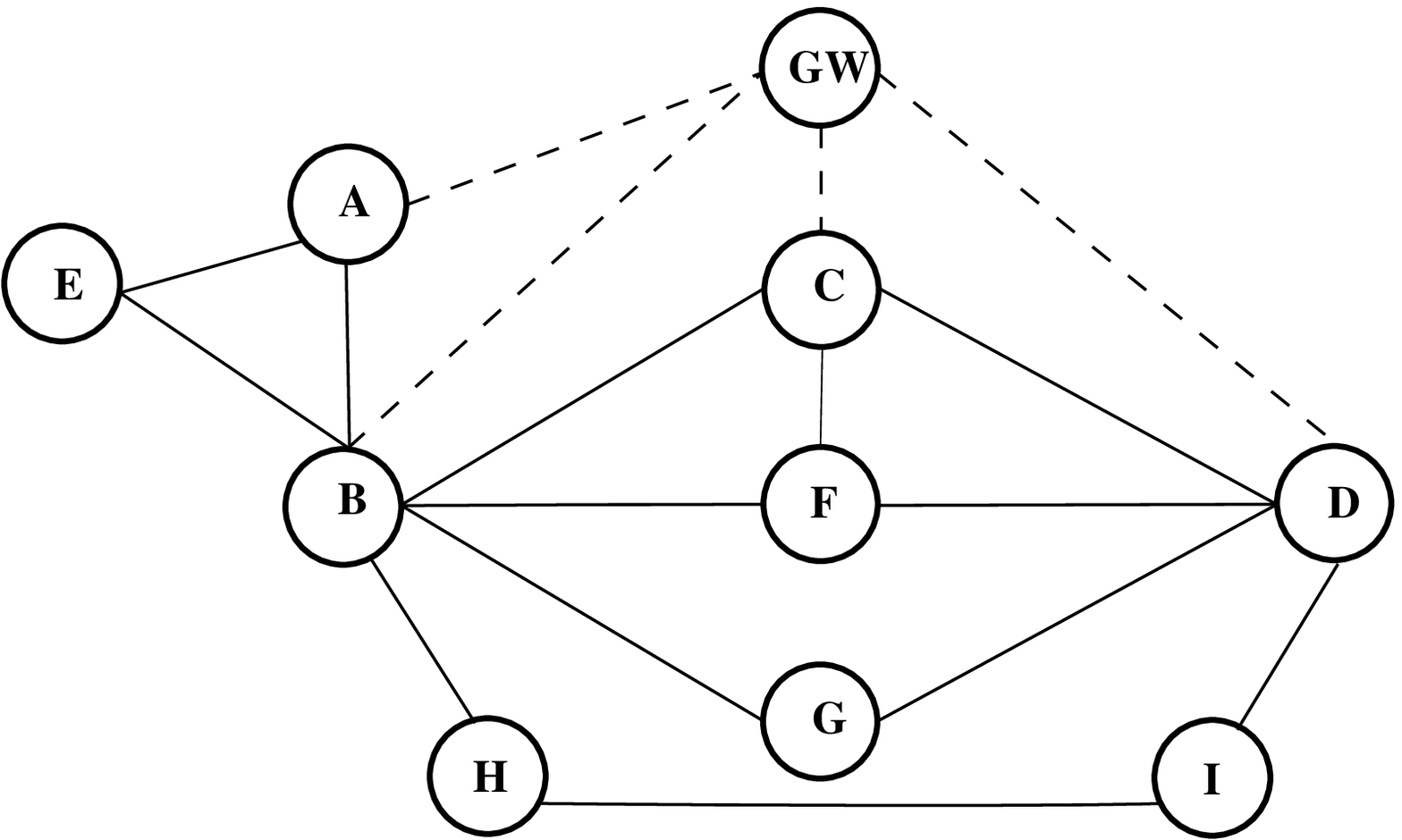}}\hfill%
   \subfloat[Sequence of EIZs] {\includegraphics[width=.5\linewidth]{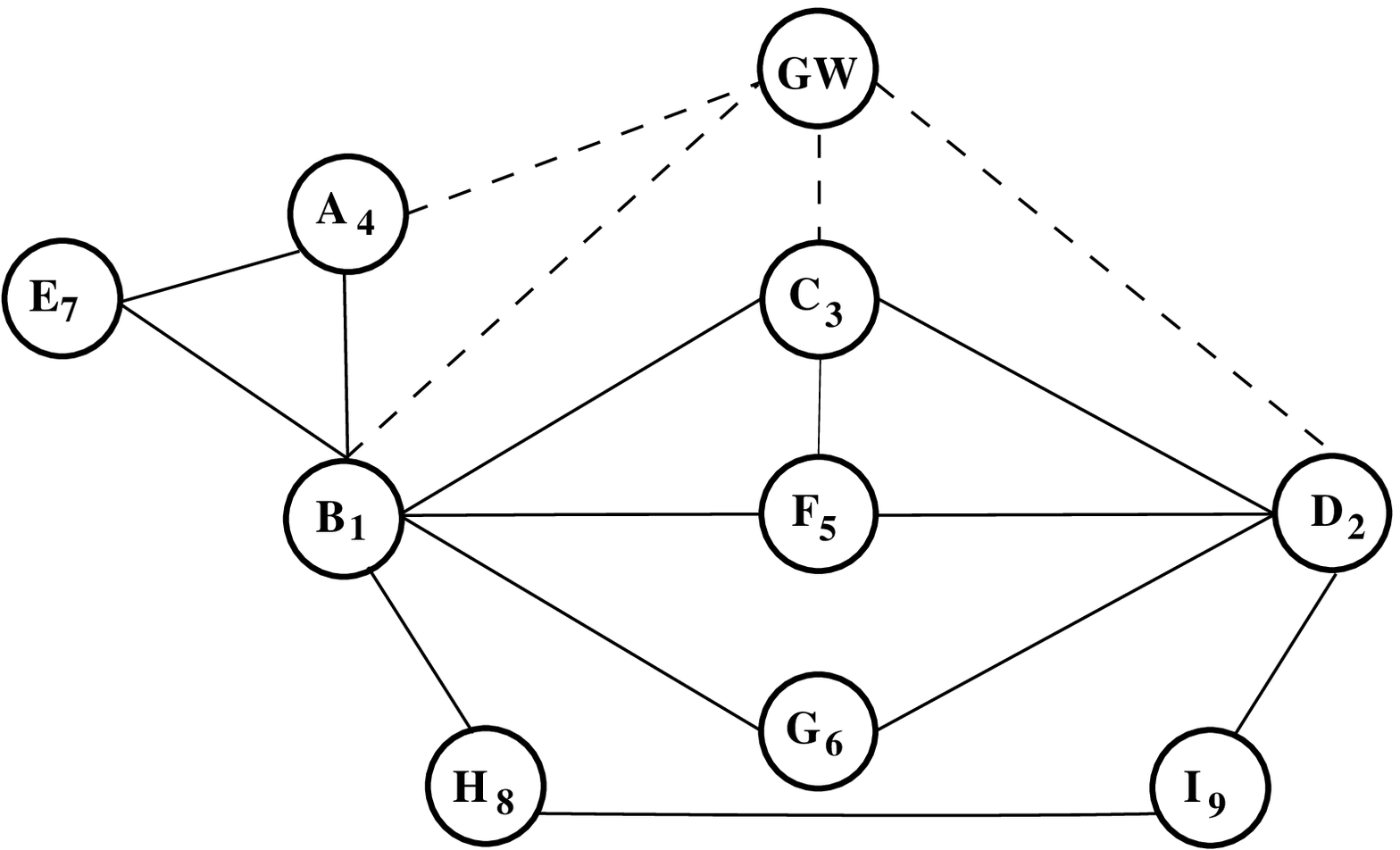}}%
    \end{tabular}
    \caption{EIZ selection in \textit{EIZM-CA} }
     \label{eizm}
\end{figure}

The next step entails processing each level-set, and assigning channels to the nodes within a level-set, iteratively. The EIZs are identified in each level-set. When a level-set is being processed for the first time, the maximal-degree node in the level-set determined by the function $MaximalDegNode$, serves as the initial EIZ node. In subsequent iterations the function $MaximalMutualNeighbor$ choses the node which boasts of the highest number of mutual neighbors with the previous EIZ node to be the current EIZ node. If there is more than one node to pick from, the node with the highest degree among them is selected as the next EIZ. Thereafter, the EIZ node is re-assigned a channel from CS which results in a minimum TID value. This step may or may not alter the initial assignment, depending upon the TID estimate. Upon final assignment, the EIZ node is removed from the level-set and helps in identifying the next EIZ node. After the initial channel allocation the algorithm ensures topology preservation through 
the steps described earlier in OIS-CA implementation. The last and the most important step is the RCI optimization proposed in Algorithm~\ref{radcol}, which has already been elaborated upon.\\
The illustrations presented in Figure~\ref{eizm} provide an insight into the functioning of the EIZM algorithm. Figure~\ref{eizm} (a) is an E-MMCG representation of a sample WMN in which the reference node signifies a corresponding link to the Gateway in $G$, and is denoted by $GW$. Based on the proposed characteristics of an EIZ, the primary potential EIZ candidates are nodes $B$ and $D$, and upto a lesser extent, nodes $C$ and $F$. This inference is substantiated by the sequence of EIZ node identification carried out by EIZM scheme, depicted in Figure~\ref{eizm} (b). The number in the subscript of the node label is the sequence number of the order in which EIZ nodes were identified. The algorithm considers $GW$ as the reference and performs a BFT, yielding {$B$, $D$, $C$, $A$} as the elements of the 1\textsuperscript{st} level-set \emph{i.e.}, the set of nodes one hop away from $GW$. Nodes of the 1\textsuperscript{st} level-set are linked to GW through dotted lines in Figure~\ref{eizm} (b). As expected, 
node $B$ is the 
1\textsuperscript{st} EIZ, followed by $C$, $D$ and $A$. The 2\textsuperscript{nd} level-set which comprises of all the remaining nodes, produces the sequence $F$, $G$, $E$, $H$ and $I$ of consecutive EIZ nodes. It may be argued that $A$ was chosen as an EIZ before $F$, despite the latter exhibiting stronger EIZ characteristics. The underlying reason is that node $A$ is one hop away from $GW$, an element of the 1\textsuperscript{st} level-set, and thus processed before $F$ which lies in the 2\textsuperscript{nd} level-set. This however, is not an anomaly but a consequence of a conscious design choice to divide the E-MMCG nodes into level-sets. The mechanism serves the objective of fair distribution of channels among radios and ensures minimal conflict between nodes of consecutive level-sets through channel initialization of each level-set in a cyclic manner. Hence, the EIZM algorithm maintains a fine balance between the spatial and statistical aspects 
of interference mitigation, and benefits by doing so.\\
The BFT approach is often used in CA schemes. For example in the BFS-CA algorithm suggested in \cite{22Ramachandran}, a BFT is performed on the multi-radio conflict graph (MCG) of the WMN, starting from a Gateway node. The nodes are accessed in the increasing order of hop-count from the Gateway, and assigned channels. However, BFS-CA and other BFT based algorithms, do not take into account the existence of high interference zones, and hence do not prioritize their channel assignment based on the concept of EIZ. Secondly, although BFS-CA processes nodes on the basis of hop-counts, it does not initialize nodes situated one hop-count away from each other with different channels. Thus, it fails to facilitate a balanced and fair channel distribution. Finally, like all other CA schemes, BFS-CA too fails to acknowledge RCI and take measures to alleviate it. Experimental results will demonstrate that EIZM scheme, designed in conformity with the proposed concepts, registers a network capacity which is around $2.5$ 
times that of BFS-CA for a few test scenarios.

\subsection{Time Complexities Of Proposed Algorithms}
For a given WMN graph $G$ of $n$ nodes and its E-MMCG $G_c$ of $m$ nodes, we derive computational costs of the proposed RCA CA schemes. Since both OIS-CA and EIZM-CA accept the E-MMCG $G_c$ as input, their time complexities are a function of $m$. In OIS-CA, generating the ISs takes $O(m\textsuperscript2)$ time, as does the BFT in EIZM-CA. In contrast, the topology preservation technique operates on WMN graph $G$ as input. It first numbers the nodes in $G$ and then for each node, traverses its adjacent nodes. Thus it has a worst-case cost of $O(n\textsuperscript2)$. Radio co-location optimization function has an $O(n\textsuperscript2)$ complexity as well. However, due to high number of conflict links in a MRMC WMN, the number of nodes in the E-MMCG are much greater than those in the original WMN \emph{i.e.,} $m>>n$. Thus, it is reasonable to conclude that the overall time complexity of both the RCA CA algorithms is  $O(m\textsuperscript2)$. 
\section{Simulations, Results and Analysis}
It is imperative that we prove the relevance of the proposed RCA CA schemes and validate the arguments their design is predicated on, through extensive experimental results. Our objectives are two-fold. 
\subsubsection{Demonstrate the significance of Radio Co-location Optimization} We highlight that RCI mitigation through RCO functionality substantially enhances a CA's performance, consolidating the primary premise of this study. This is accomplished by comparing the instance of the proposed CA schemes that is non-RCA \emph{i.e.}, the CA generated before RCO step, with the corresponding final RCA CA. The RCA CAs are denoted by \textit{OIS-CA \& EIZM-CA} while their non-RCA counterparts are represented by \textit{OIS-N-CA \& EIZM-N-CA} in the result illustrations. 
\subsubsection{Compare performance of RCA CAs with conventional CAs} We present experimental evidence to corroborate that RCA CAs significantly outperform the conventional CAs that do not ensure RCI alleviation and lack a spatio-statistical design. We employ MaIS-CA \cite{24Aizaz} and BFS-CA \cite{22Ramachandran}, as the reference CAs against which we compare OIS-CA and EIZM-CA, respectively. The reference CAs are carefully chosen to demonstrate that despite a likeness in the underlying approach of the CA pairs OIS-CA \& MaIS-CA, and EIZM-CA \& BFS-CA, elaborated upon earlier, the RCA algorithms fare remarkably better.


\subsection{Simulation Parameters}

We perform exhaustive simulations in ns-3 \cite{NS-3} to gauge the performance of CAs deployed on the following two WMN topologies.
\begin{itemize}
 \item $5\times5$ grid WMN depicted in Figure~\ref{TID} (b).
 \item A random WMN of 50 nodes spread across an area of $1500m \times 1500m$.
\end{itemize}

The choice of a grid WMN is motivated by the fact that they fare better than random layouts in terms of coverage area and mesh network capacity \cite{Grid}, resulting in an ideal topology for CA performance evaluation. The simulated environment of a large WMN comprising of 50 randomly placed nodes is also a relevant layout, as it resembles a real-world WMN deployment which is less likely to conform to a grid pattern. The large number of nodes spread over a wide area facilitates long distance data transmissions requiring multiple-hops between nodes placed on the fringes of the network, a scenario vital for estimating the efficiency of a CA in performing such transmissions.
For ease of reference, grid WMN and random WMN are abbreviated as \textit{GWMN} and \textit{RWMN}, respectively.
The simulation parameters are presented in Table~\ref{Sim}. Each multi-hop traffic flow transmits a datafile from the source to the destination. 
We carry out two set of simulations employing TCP and UDP as the underlying transport layer protocols. We leverage the inbuilt ns-3 models of \textit{BulkSendApplication} and \textit{UdpClientServer} for TCP and UDP implementations, respectively. TCP simulations are aimed at estimating the \textit{Aggregate Network Throughput}, which we henceforth simply refer to as the \textit{Throughput$_{Net}$}, for each scenario. UDP simulations are employed to determine the \textit{packet loss ratio} (PLR) and the \textit{mean delay} (MD) for a test-scenario.
 \begin{table} [h!]
\caption{ns-3 Simulation Parameters}
\raggedright
\begin{tabular}{|M{4.5cm}|M{3.5cm}|}
\hline
\bfseries
 Parameter&\bfseries Value \\ [0.2ex]
 \hline
\hline
Radios/Node&GWMN: 2,    RWMN: 3\\
\hline
Range Of Radios&250 mts   \\
\hline
IEEE Protocol Standard &GWMN: 802.11g      RWMN: 802.11n  \\
\hline
Available Orthogonal Channels&GWMN: 3 (2.4 Ghz)  RWMN: 4 (5 Ghz)\\
\hline
Transmitted File Size &GWMN: 10 MB      RWMN: 1 MB  \\
\hline
Maximum 802.11g/n Phy Datarate&54 Mbps  \\
\hline
Maximum Segment Size (TCP)&1 KB   \\
\hline
Packet Size (UDP)     &GWMN: 1 KB \quad    RWMN: 512 Bytes\\
\hline
MAC Fragmentation Threshold&2200 Bytes  \\
\hline
RTS/CTS&Enabled  \\
\hline
Packet Interval (UDP) & 50ms  \\
\hline
Routing Protocol Used &OLSR    \\
\hline
Loss Model&Range Propagation   \\
\hline
Rate Control&Constant Rate   \\
\hline
\end{tabular}
\label{Sim}
\end{table}   
\subsection{Data Traffic Characteristics}
Tailoring an ideal set of \textit{data traffic characteristics} is a crucial step so as to aptly highlight the performance bottlenecks caused by the endemic interference and to explicitly demonstrate the mitigation of their adverse impact on the WMN performance by the deployed CA. 
Since, multi-hop data flows are an inherent feature of WMNs, we simulate a variety of test scenarios for both the WMN topologies which employ the following multi-hop flows.
\subsubsection {Grid WMN} 4-Hop Flows or 4HFs are established from the first node (source) to the last node (sink), of each row and each column of the grid. 8-Hop Flows or 8HFs are set up between the diagonal nodes of the grid.  
\subsubsection {Random WMN} We create a plethora of multi-hop flows between nodes which are $3$ to $10$ hops away, ensuring that the paths of many of these flows intersect to occasion comprehensive interference scenarios.

\subsection{Test Scenarios}
Various combinations of the multi-hop flows described above are devised to generate test-scenarios that reflect both, a \textit{sectional} view and a \textit{comprehensive} view, of the intensity, impact and alleviation of the prevalent interference. Test-cases for both the WMN layouts are listed below.
\subsubsection {Grid WMN}
\begin{enumerate}[label=(\alph*)]
 \item \textit{D2} : Both 8HFs  concurrently.
 \item \textit{H5} : All five 4HFs concurrently in the horizontal direction. 
 \item \textit{V5} : All five 4HFs concurrently in the vertical direction. 
 \item \textit{H4V4} : Variety of eight concurrent flows, which include various combinations of 4HFs.
 \item \textit{H5V5} : Ten concurrent 4HFs \emph{viz.} H5 \& V5.
 \item \textit{H5V5D2} : Twelve concurrent flows. \emph{viz.} D2, H5 \& V5.
\end{enumerate}
Test cases \textit{(a,b \& c)} offer a directional perspective of CA performance, while scenarios \textit{(d, e \& f)} form the exhaustive benchmarks on which the overall performance of a CA can be assessed.  
\subsubsection {Random WMN}
Given the random placement of nodes, the test-scenarios include a combination of concurrent multi-hop flows of varying hop counts. 4, 8, 12, 16 and 20 concurrent multi-hop flows were established, where the number of simultaneous data flows represents a test-case \emph{viz.} \textit{TC4, TC8, TC12, TC16 \& TC20}. As the number of concurrent flows increases, the interference dynamics become more complex. Thus test-cases TC12, TC16 and TC20 are ideal to gauge CA efficiency in terms of network performance metrics.
\begin{figure}[htb!]
                \includegraphics[width=8cm, height=5cm]{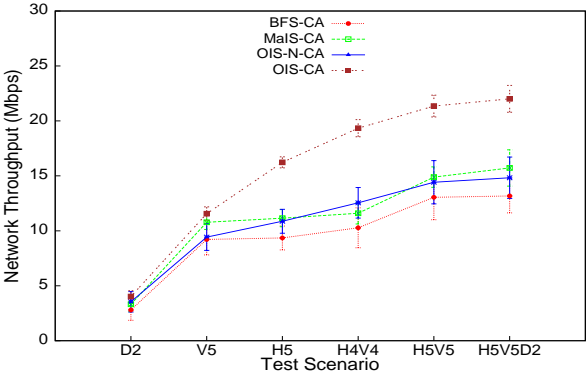}
                \caption{GWMN Throughput$_{Net}$  : \textit{OIS CA}}
                \label{OISGT}
                	 \vspace{.1cm}
        \end{figure}
\begin{figure}[htb!]
\includegraphics[width=8cm, height=5cm]{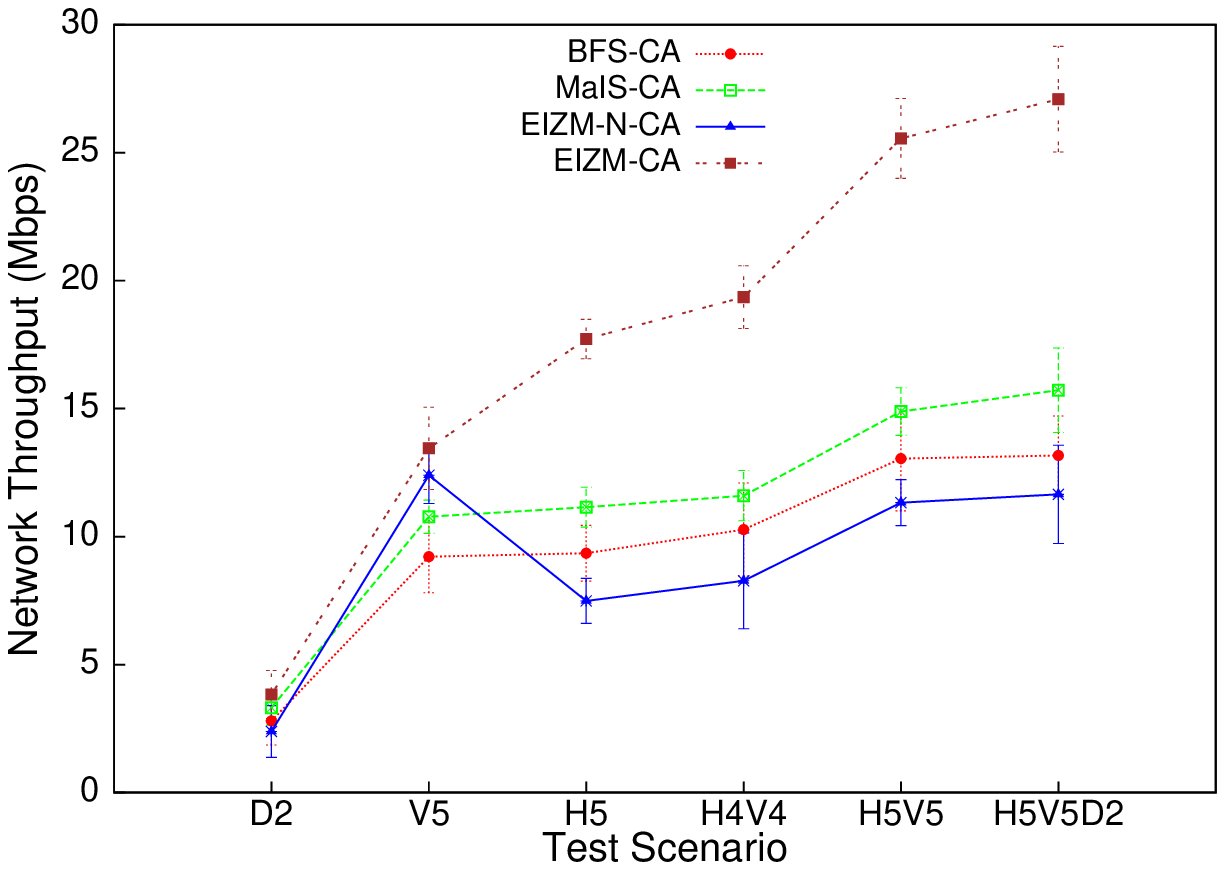}
\caption{GWMN Throughput$_{Net}$  : \textit{EIZM CA}}
\label{EIZMGT}
	  \vspace{.1cm}
\end{figure}
\begin{figure}[htb!]
                \includegraphics[width=8cm, height=5cm]{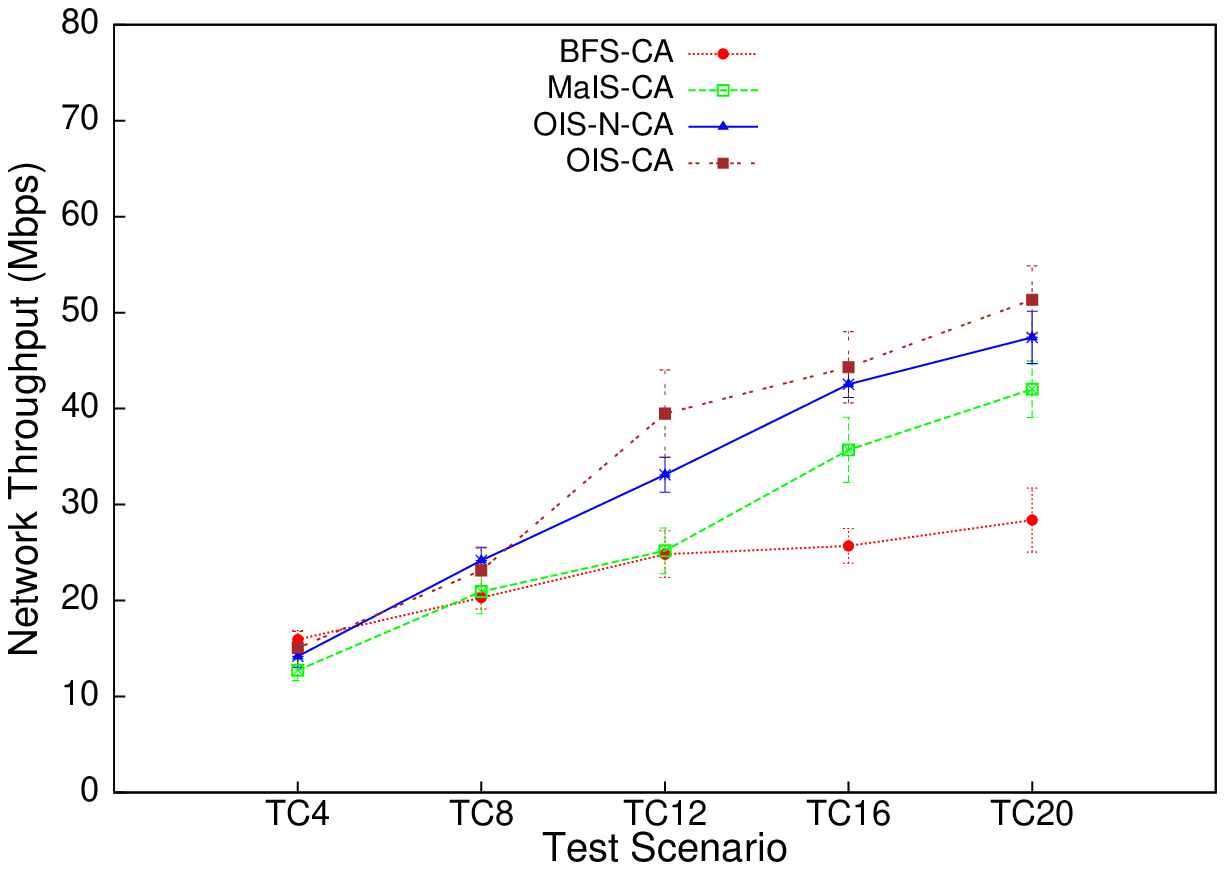}
                \caption{RWMN Throughput$_{Net}$  : \textit{OIS CA}}
                \label{OISRT}
                	 \vspace{.1cm}
        \end{figure}
\begin{figure}[htb!]
	\includegraphics[width=8cm, height=5cm]{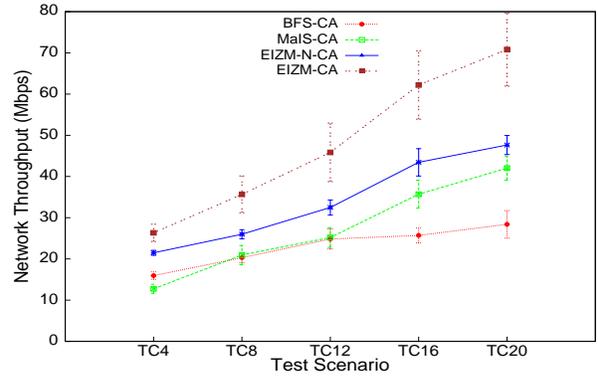}
	\caption{RWMN Throughput$_{Net}$  : \textit{EIZM CA}}
	\label{EIZMRT}
		 \vspace{.1cm}
\end{figure}

\subsection{Results and Analysis}
Rigorous simulations were run for the test-cases described above, and the observed values of performance metrics \emph{viz.} Throughput$_{Net}$ (Mbps), PLR (\% of packets lost) and MD ($\mu$seconds) are now presented for a thorough analysis of the performance of the proposed RCA CAs. 
\subsubsection{Throughput$_{Net}$}
The throughput results for simulations run on GWMN and RWMN topologies are depicted in Figures~\ref{OISGT}, \ref{EIZMGT}, \ref{OISRT} \& \ref{EIZMRT}. For statistical reliability \textit{99\% Confidence Interval} bars have been marked for the recorded Throughput$_{Net}$ value of each test-case. Our first objective is met through a conspicuous inference from the listed plots, that the non-RCA version of the proposed CAs registers much lower throughput than the RCA version, highlighting the efficacy of radio co-location optimization feature of the RCA CAs and consolidating our theoretical contention that RCI mitigation enhances the network capacity significantly. EIZM-CA registers a maximum capacity enhancement of $132\%$ for scenario H5V5D2 in GWMN and about $48\%$ for scenario TC20 in RWMN, over EIZM-N-CA. Improvements exhibited by OIS-CA over OIS-N-CA for these scenarios are $48\%$ and $8\%$, respectively. Let us now analyze the performance of RCA CAs in comparison to reference CAs. RCA CAs outperform both 
MaIS-CA and BFS-CA by a significant margin, which is evident from the substantial increase in Throughput$_{Net}$ in RCA CA deployments, displayed in \mbox{Table~\ref{T1}}. 

\begin{table} [h!]
\raggedright
\center
\caption{Enhancement in network capacity through RCA CAs}
\tabcolsep=0.11cm
\begin{tabular}{|M{3.4cm}|M{1cm}|M{1cm}|M{1.1cm}|M{1.3cm}|}
\hline 
    \multicolumn{1}{|c}{} & \multicolumn{4}{|c|}{\textbf{\% increase in Throughput$_{Net}$}}\\ 
    \multicolumn{1}{|c}{} & \multicolumn{4}{|c|}{\textbf{in TC}}\\  \cline{2-5}
    \multicolumn{1}{|c|}{\textbf{Comparing CAs }}&\textbf{TC16}&\textbf{TC20}&\textbf{H5V5}&\textbf{H5V5D2}\\
\hline  
EIZM-CA vs BFS-CA&142&149&96&106\\
\hline  
EIZM-CA vs MaIS-CA&74&68&72&72\\
\hline  
OIS-CA  vs BFS-CA&72&81&64&67\\
\hline  
OIS-CA  vs MaIS-CA&24&22&43&40\\
\hline  

\end{tabular} 
\label{T1}
\end{table}

EIZM-CA turns out to be the better of the two high performance RCA CAs, however OIS-CA fares quite better than both BFS-CA and MaIS-CA as well. Another interesting observation is that the non-RCA CAs perform decidedly better than both of the reference schemes in the RWMN simulations, but not in the GWMN. Hence, no definitive conclusions can be made between non-RCA CAs and the the reference CAs, which highlights the importance of RCI alleviation in enhancing CA performance. Further, OIS-N-CA performs slightly better than OIS-CA in test-case TC8 in RWMN. But this scenario projects a partial or sectional view of the RWMN and the upset in results does not amount to a reversal in the expected trend. This momentary aberration is remedied in the remaining test-cases, where OIS-CA continues to outperform OIS-N-CA. Another noteworthy point is the spatial impact of interference in GWMN where the scenarios H5 and V5 include an equal number of (five) concurrent 4HFs, along the rows and columns of the grid, 
respectively. Yet in Figure~\ref{EIZMGT} it can be observed that for EIZM-CA, scenario H5 records a higher throughput than V5 \emph{i.e.}, lower impact of interference is experienced by transmissions along the rows, while for EIZM-N-CA the situation is reversed \emph{i.e.}, transmissions along the columns register higher throughput.
\subsubsection{Packet Loss Ratio}
\begin{figure}
  \centering%
  \begin{tabular}{cc}
   \subfloat[PLR of OIS-CA]{\includegraphics[width=.5\linewidth]{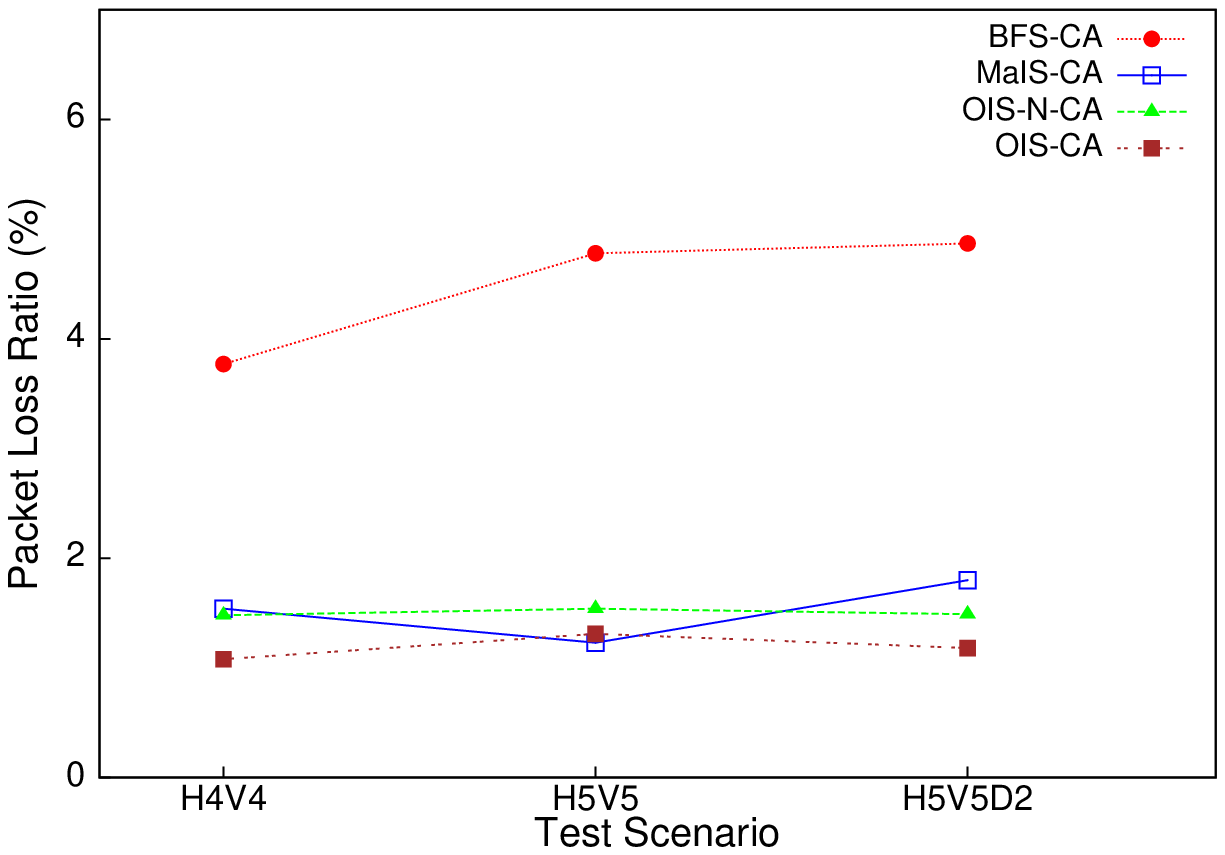}}\hfill%
   \subfloat[PLR of EIZM-CA] {\includegraphics[width=.5\linewidth]{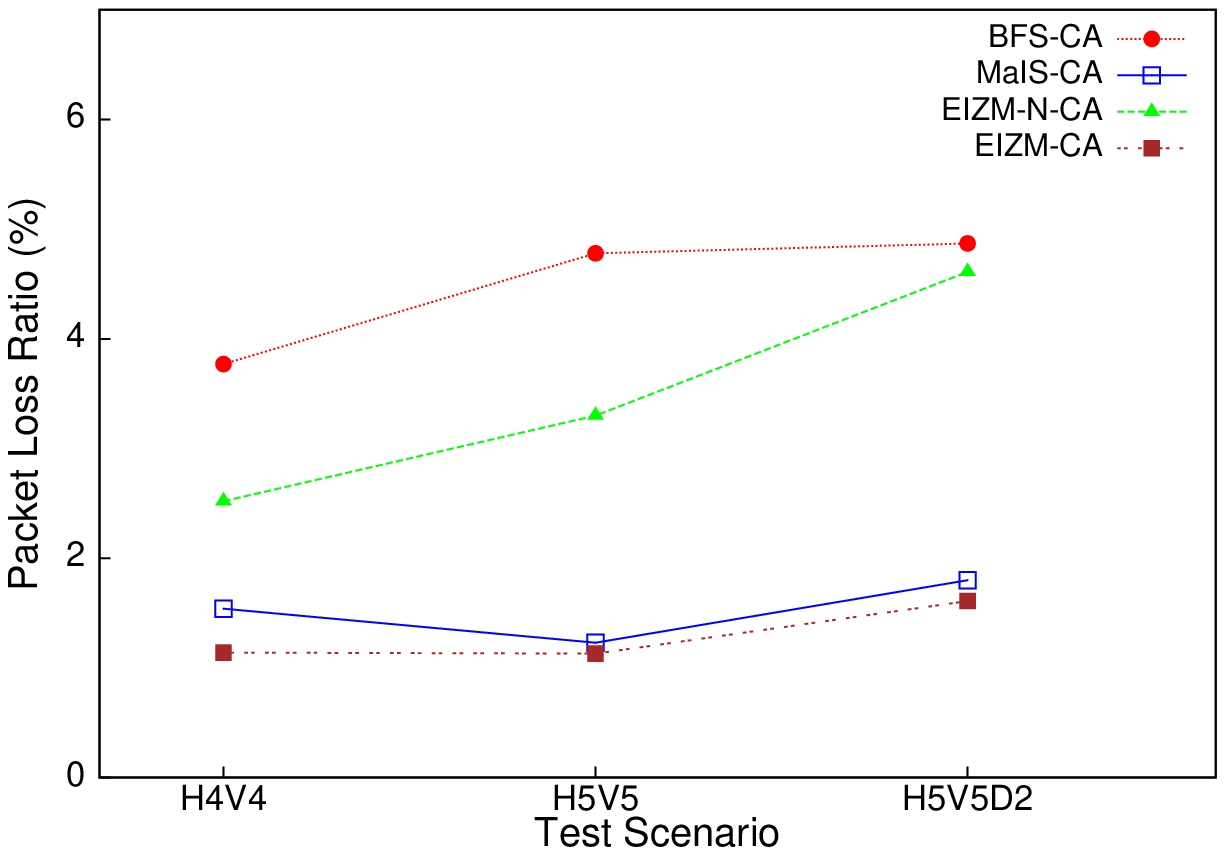}}%
    \end{tabular}
    \caption{PLR of RCA CAs in GWMN} 
     \label{gPLR}
\end{figure}

\begin{figure}
  \centering%
  \begin{tabular}{cc}
   \subfloat[PLR of OIS-CA]{\includegraphics[width=.5\linewidth]{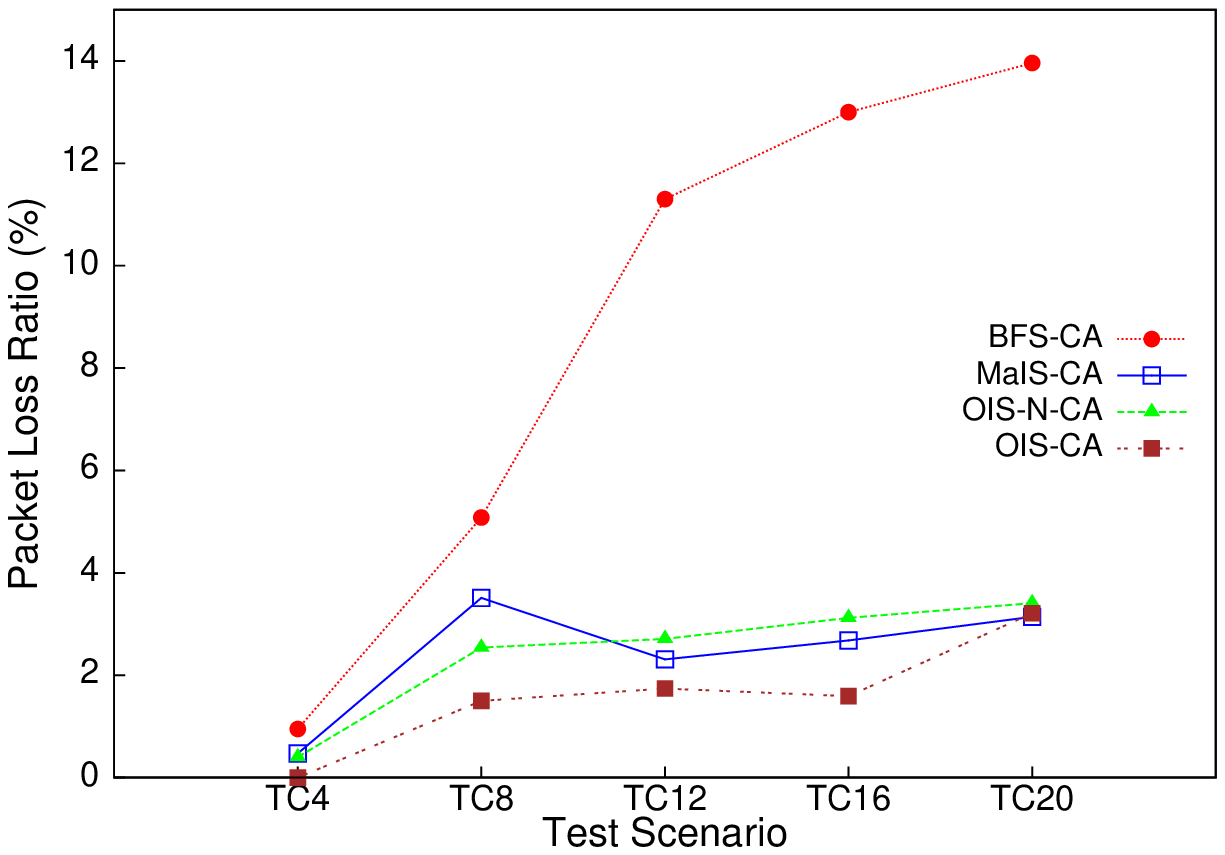}}\hfill%
   \subfloat[PLR of EIZM-CA] {\includegraphics[width=.5\linewidth]{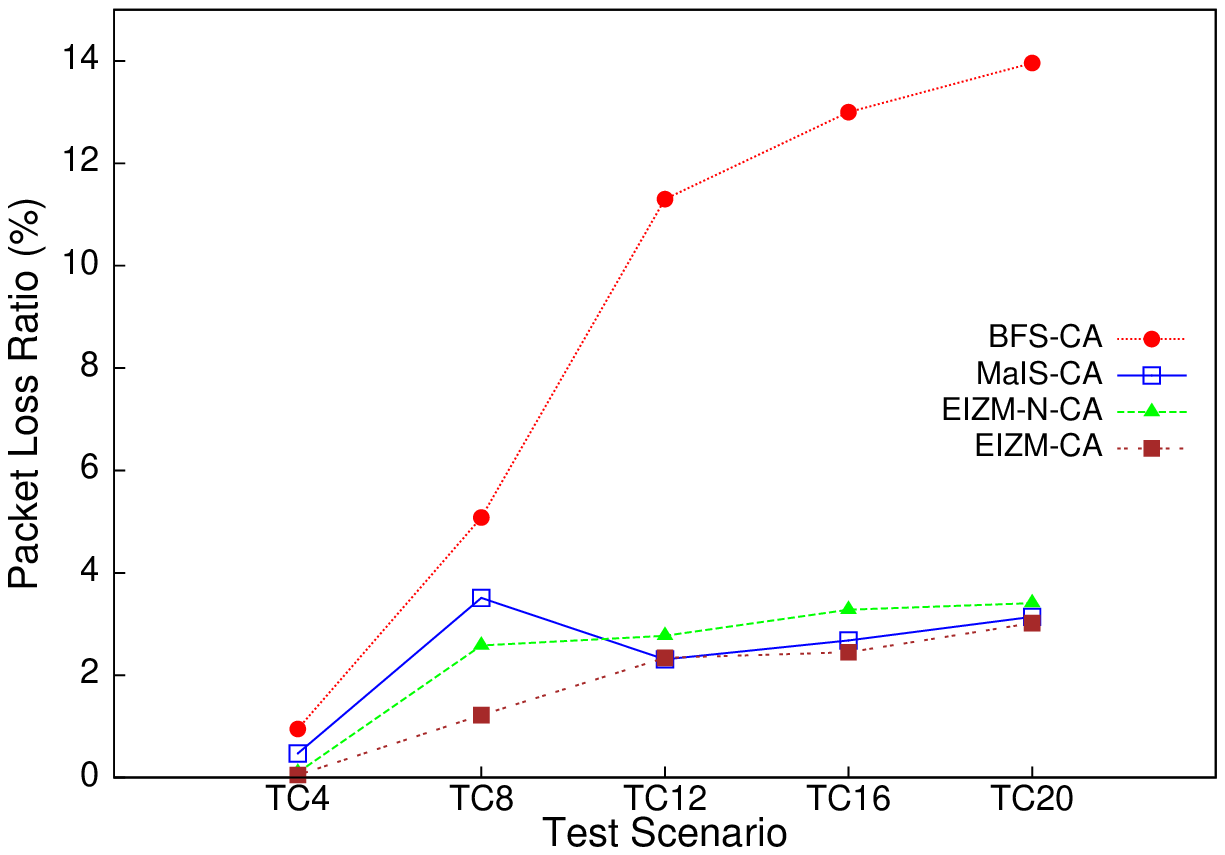}}%
    \end{tabular}
    \caption{PLR of RCA CAs in RWMN} 
     \label{rPLR}
\end{figure}

PLR performance metric observations are in absolute conformity with the network capacity results. For the RWMN layout, we present PLR for all the test-cases, however for the GWMN topology, results of the three comprehensive test-cases \emph{viz.}, H4V4, H5V5 and H5V5D2, are presented as the PLR in other scenarios was negligible. In Figures~\ref{gPLR} and \ref{rPLR}, it can be noticed that both the RCA CAs experience minimum packet loss. 
Thus, PLR estimates also highlight the enhancement in performance of deployed RCA CAS  owing to their twin capabilities of RCI alleviation and spatio-statistical design. For a quantitative analysis, the \% reduction in PLR effected by RCA CAs in comparison to the reference CA approaches is elicited in \mbox{Table~\ref{T2}}.

\begin{table} [h!]
\caption{Reduction in PLR through RCA CAs}
\tabcolsep=0.11cm
\begin{tabular}{|M{3.4cm}|M{1cm}|M{1cm}|M{1.1cm}|M{1.2cm}|}
\hline 
    \multicolumn{1}{|c|}{} & \multicolumn{4}{|c|}{\textbf{\% decrease in PLR in TC}}\\  \cline{2-5}
    \multicolumn{1}{|c|}{\textbf{Comparing CAs }}&\textbf{TC16}&\textbf{TC20}&\textbf{H5V5}&\textbf{H5V5D2}\\
\hline  
EIZM-CA vs BFS-CA&81&78&76&67\\
\hline  
EIZM-CA vs MaIS-CA&9&4&8&11\\
\hline  
OIS-CA  vs BFS-CA&88&77&73&76\\
\hline  
OIS-CA  vs MaIS-CA&41&-2&-6.5&34\\
\hline  
\end{tabular} 
\label{T2}
\end{table}

There are three minor upsets in the observed trends, two of which are between OIS-CA and MaIS-CA, where the latter performs marginally better in test cases TC20 and H5V5. But these slight reversals are mere exceptions in the observed pattern, and do little to discredit the improvement exhibited by RCA CAs. Further, the two RCA CAs exhibit a similar performance in terms of packets lost during transmission.

\begin{figure}
  \centering%
  \begin{tabular}{cc}
   \subfloat[MD of OIS-CA]{\includegraphics[width=.5\linewidth]{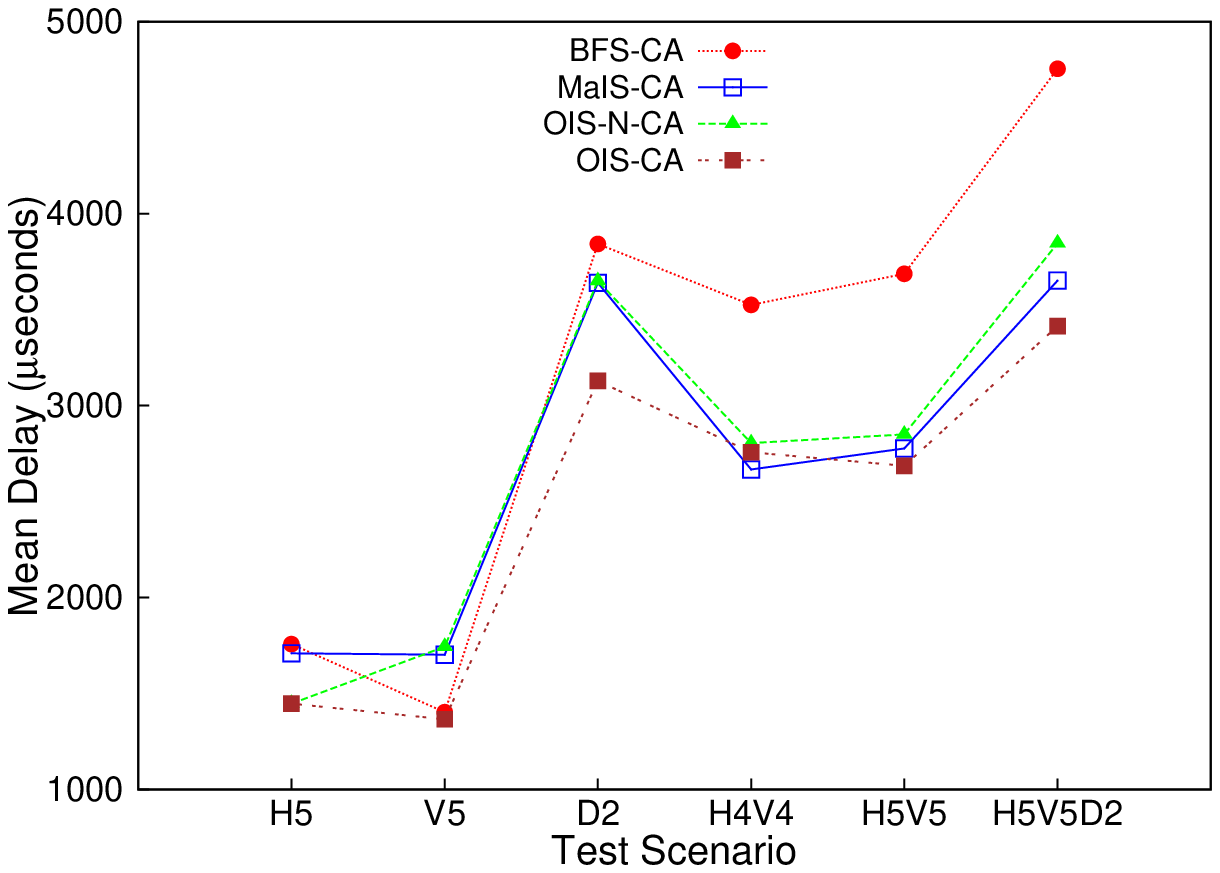}}\hfill%
   \subfloat[MD of EIZM-CA] {\includegraphics[width=.5\linewidth]{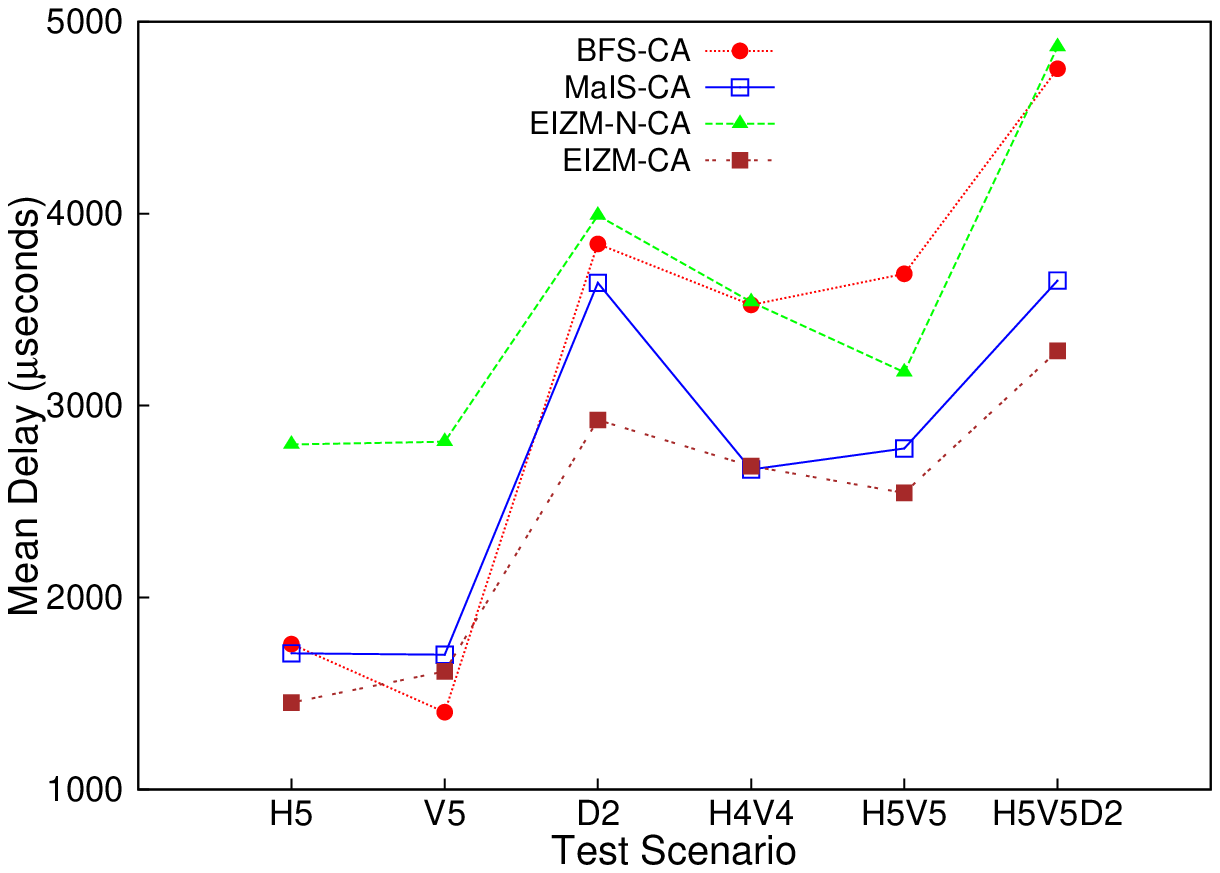}}%
    \end{tabular}
    \caption{MD of RCA CAs in GWMN} 
     \label{gMD}
\end{figure}

\begin{figure}
  \centering%
  \begin{tabular}{cc}
   \subfloat[MD of OIS-CA]{\includegraphics[width=.5\linewidth]{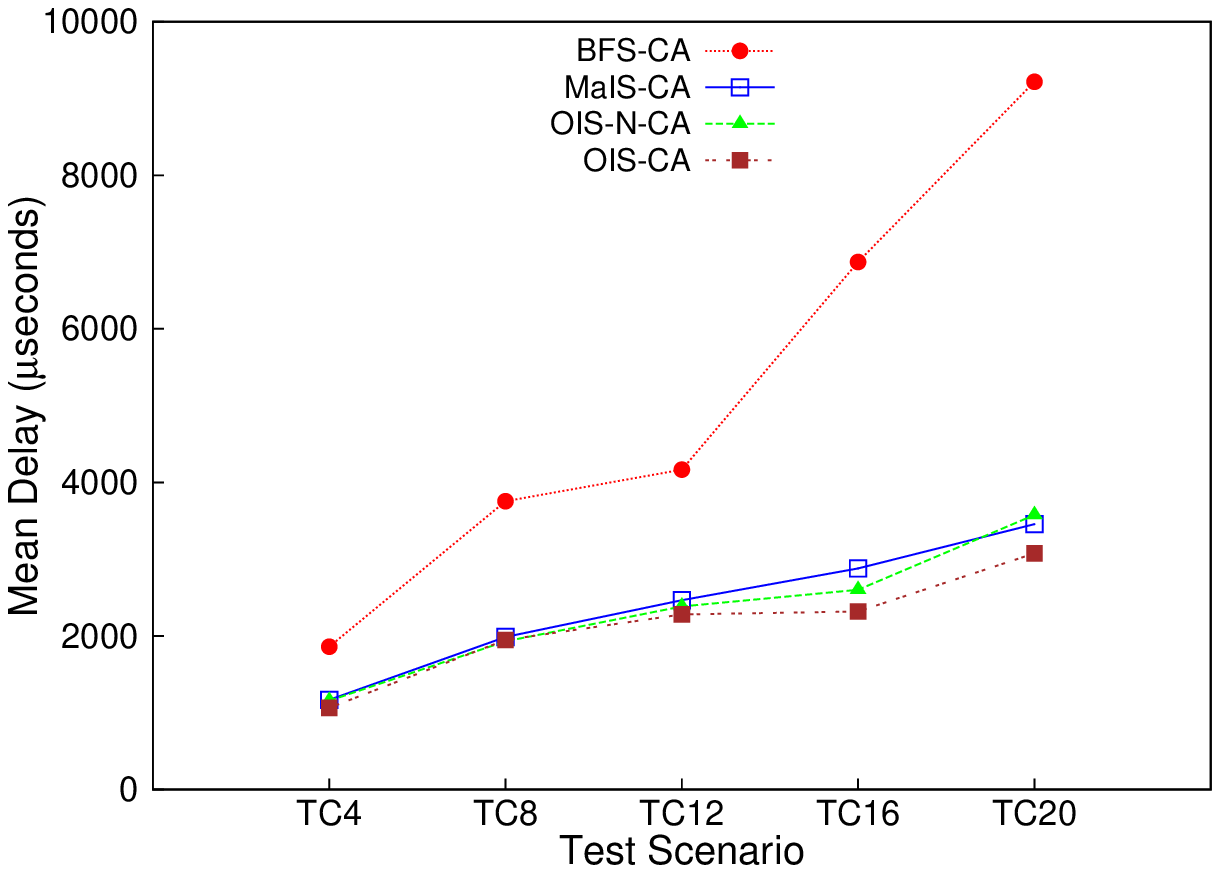}}\hfill%
   \subfloat[MD of EIZM-CA] {\includegraphics[width=.5\linewidth]{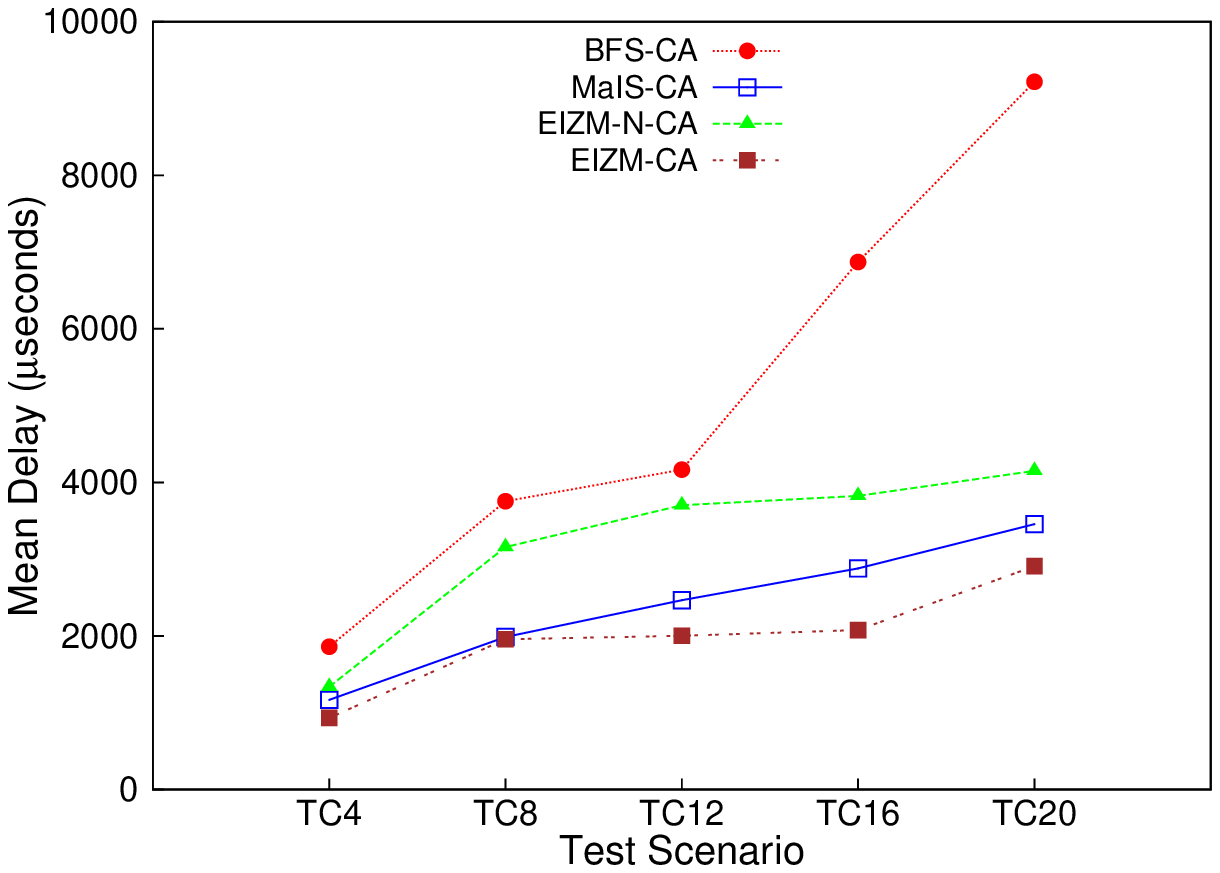}}%
    \end{tabular}
    \caption{MD of RCA CAs in RWMN} 
     \label{rMD}
\end{figure}
\subsubsection{Mean Delay}
Multi-hop flows are an inherent characteristic of the WMNs. It is of great relevance to observe the ease with which, for a CA implementation, data is transmitted across distant nodes in the WMN that are numerous hop-counts apart, especially under high network traffic loads. The recorded MD values are illustrated in Figures~\ref{gMD} and \ref{rMD}, for simulations run on GWMN and RWMN topologies, respectively. The delay characteristics of all the CAs are similar to those observed in PLR results. RCA CAs reduce the MD time considerably, especially in comparison to BFS-CA. 
This observation is in adherence to the notion that low PLR generally leads to small packet delays. There are no noticeable reversals and RCA CAs consistently perform better than the reference CAs. Further, between the two RCA CAs, none has a distinct  edge over the other. While OIS-CA exhibits small delay times for some test-
cases, EIZM-CA registers a greater reduction in MD for others.

\begin{table} [h!]
\caption{Reduction in MD through RCA CAs}
\tabcolsep=0.11cm
\begin{tabular}{|M{3.4cm}|M{1cm}|M{1cm}|M{1.1cm}|M{1.2cm}|}
\hline 
    \multicolumn{1}{|c|}{} & \multicolumn{4}{|c|}{\textbf{\% decrease in MD in TC}}\\  \cline{2-5}
    \multicolumn{1}{|c|}{\textbf{Comparing CAs }}&\textbf{TC16}&\textbf{TC20}&\textbf{H5V5}&\textbf{H5V5D2}\\
\hline  
EIZM-CA vs BFS-CA&70&68&31&31\\
\hline  
EIZM-CA vs MaIS-CA&28&16&8&10\\
\hline  
OIS-CA  vs BFS-CA&66&67&27&28\\
\hline  
OIS-CA  vs MaIS-CA&19&11&3&6\\
\hline  
\end{tabular} 
\label{T3}
\end{table}
These findings provide an experimental validation to the emphasis we have laid on the spatio-statistical design of CA algorithms which is further improved through RCI alleviation measures.

\section{Conclusions}
Having thoroughly deliberated over the performance of RCA CAs, we now make some sound logical conclusions. First, RCI prevalent in a WMN has a significant adverse impact on the network capacity, and its mitigation leads to enhanced Throughput$_{Net}$ and reduced PLR in a wireless network. Thus RCI mitigation ought to be a primary consideration in a CA scheme. Second, a prudent CA design that is tuned to the spatio-statistical aspects of interference alleviation will be more effective in restraining the detrimental effects of interference. Further, a CA scheme which caters to both spatial and statistical dimensions, such as EIZM-CA, stands to fare quite better than one which considers only one of the aspects in its design, such as OIS-CA. On the subject of the performance of the proposed RCA CAs, it can be unarguably concluded that they are significantly better than the reference CA schemes in terms of Throughput$_{Net}$, PLR and MD. They are high-performance CAs which enhance network capacity tremendously, 
are 
resilient to packet loss, and reduce data propagation delays. 

Considering the average of network metrics, EIZM-CA outperforms OIS-CA in terms of network capacity, by $45\%$ and $11\%$ in RWMN and GWMN layouts, respectively. The two RCA CAs are at par with respect to packet mean delay and PLR. The better performance of OIS-CA in GWMN can be attributed to the structured layout of a grid which facilitates a proportional distribution of channels across radios, unlike RWMN which is a large network of randomly placed nodes. EIZM-CA registers high performance in both WMN topologies, regardless of the layout. Clearly, EIZM-CA is the more efficient and better performing CA algorithm of the two, which it owes to its spatio-statistical design coupled with the feature of RCI mitigation.
%
%

%

\bibliography{WoWMoM}

\end{document}